\documentstyle[preprint,aps,epsf]{revtex}

\newcommand{\be}{\begin{eqnarray}}
\newcommand{\ee}{\end{eqnarray}}

\begin{document}
\draft
\title{Hadronic Correlation Functions \\
       in the Interacting Instanton Liquid}

\author{T.~Sch\"afer}

\address{Institute for Nuclear Theory, Department of Physics,
         University of Washington,\\ Seattle, WA 98195, USA}

\author{E.V.~Shuryak}

\address{Department of Physics, State University of New York at Stony
         Brook,\\ Stony Brook, NY 11794, USA}

\maketitle

\begin{abstract}
    In this paper we study hadronic correlation functions in the
interacting instanton liquid model, both at zero and nonzero
temperature $T$. At zero $T$ we investigate the dependence of
the correlators on the instanton ensemble, in particular the
effect of the fermionic determinant. We demonstrate that
quark-induced correlations between instantons are important,
especially in the repulsive $\eta'$ and $\delta$-meson channels.
We also calculate a large number of mesonic and baryonic correlation
functions as a function of temperature. We find three
different types of behavior as $T\to T_c$. The vector channels
$\rho,a_1,\Delta$ show a gradual melting of the resonance
contribution and approach free quark behavior near the chiral
phase transition. The light pseudoscalars and scalars $\pi,\sigma$,
as well as the nucleon show stable resonance contributions,
probably even surviving above $T_c$. Correlation functions in the
heavy scalar channels $\eta',\delta$ are enhanced as $T\to T_c$.
\end{abstract}
\pacs{11.30.Rd, 12.38.Lg, 12.38.Mh}

\section{Introduction}

   Hadronic correlation functions provide a rich source of information
about the spectrum of hadronic resonances and the interaction of quarks
\cite{Shu_93}. In this work we study the behavior of point-to-point
QCD correlation functions $\Pi_h(x)= <j_h(x)j_h(0)>$. Here, $j_h(x)$
is a current with the quantum numbers of some hadron $h$, and the
averaging is performed over the QCD vacuum. For large euclidean
separations $\tau=\sqrt{-x^2}$, the correlation function $\Pi_h(\tau)$
decays exponentially, the decay being controlled by the mass and the
coupling constant of the lowest resonance in that channel. At small
distances, asymptotic freedom implies that the correlator approaches
free field behavior. The functional form at intermediate distances
can be compared with the operator product expansion and reveals
interesting information about the structure of the vacuum and the
interactions between quarks.

     In the present work we evaluate a large number of hadronic
correlation functions, both at zero and non-zero temperature, in the
interacting instanton liquid model (IILM) of the QCD vacuum. This
study is a continuation of our earlier work on the instanton model
\cite{Shu_82a,Shu_88,SV_93b,SSV_94,SS_95c}, and we refer to these
references, in particular \cite{SS_95c}, for a detailed description
of the model, its underlying assumptions, and its theoretical and
phenomenological foundations. The purpose of our study of hadronic
correlation functions is twofold. First, we want to compare the
correlation functions in the interacting ensemble with our results
for the random one \cite{SV_93b,SSV_94}. In the random model, the
collective coordinates of the instantons are distributed randomly
(except for their size, which is kept fixed), while in the interacting
model, configurations are generated according to the correct measure,
consisting of the bosonic contribution $\exp(-S)$ and, most important,
the fermion determinant $\det(\hat D+m)$. This comparison is analogous
to comparing ``quenched" and ``unquenched" lattice simulations
(``quenching" means that the fermion determinant was neglected).
The physical meaning of the quenched approximation and the role
of quark induced correlations is an important problem \cite{DT_95,LM_95}.
We will argue that the fermion determinant indeed plays a very significant
role, in particular in the pseudoscalar isosinglet ($\eta'$-meson) and
scalar isovector ($\delta$-meson) channels.

   The second purpose of our work is to study the behavior of hadronic
correlation functions at finite temperature. This study is clearly
of great interest for the behavior of hadrons under extreme conditions,
in particular near the chiral phase transition. In this regime there
is little phenomenological information and lattice simulations have
not yet progressed to the point that conclusive results can be obtained.
It is therefore important to at least gain a qualitative understanding
of the behavior of hadronic correlators near the phase transition. At
finite temperature the role of the fermion determinant becomes even
more significant. As discussed in \cite{SS_95c}, quark induced
correlations drive the chiral phase transition in the instanton
liquid. The mechanism for this transition is a rearrangement of
the instanton liquid, going from a disordered instanton liquid
to a phase of correlated instanton antiinstanton molecules.

    Finally, we perform an exploratory study of correlators in QCD
with many light flavors. QCD like theories with different matter
content form a rich (and largely unexplored) field. In the instanton
model, chiral symmetry is restored at large $N_f$. In order to
improve our understanding of this unusual phase, we would like
to study the hadronic spectrum in the large $N_f$ instanton liquid.

   The paper is organized as follows. In section 2 we calculate the
hadronic correlation functions at zero temperature, study the
dependence on the instanton ensemble and compare with phenomenological
and lattice information. In section 3 we extend our model to finite
temperature, considering correlators in the imaginary time direction.
In section 4 we evaluate correlation functions in the spatial
direction, related to the spectrum of screening masses at finite
temperature. In section 5 we do a similar study for QCD with many
light quark flavors. Our conclusions are summarized in section 6.

\section{Correlation functions at zero temperatures}

   Before we proceed to the instanton model, let us mention other
sources of information about the behavior of hadronic correlation
functions at zero temperature. General results as well as the
available phenomenological information is summarized in the review
\cite{Shu_93}. Let us again emphasize that point-to-point correlation
functions are directly related to the spectrum of physical excitations,
and any model describing hadronic states and interquark interactions
should be tested against these results. Lattice measurements of
point-to-point correlation functions were reported in
\cite{CGHN_93a,CGHN_93b,Lei_95a,Lei_95b}, while our previous results
based on the random ensemble (RILM) can be found in \cite{SV_93b,SSV_94}.
An analytical approach to the random ensemble was pioneered by Diakonov
and Petrov \cite{DP_86}, and the corresponding coordinate space
correlation functions were recently reported in \cite{Hut_95}.

   Comparing these works one should keep in mind that both
the lattice results and the random instanton model are ``quenched"
calculations: they ignore quark-induced effects in the ensemble.
On the other hand, lattice measurements of course contain a number
of effects not accounted for in the instanton model, in particular
confinement forces and perturbative interactions between quarks.
Using a method called ``cooling" one can also eliminate these effects
on the lattice, while preserving instanton contributions. The
correlation functions in the cooled ensemble were studied in
\cite{CGHN_94}.

   The correlation function are evaluated using the quark propagator
in a given configuration. For an isovector meson current $j_\Gamma=
\bar u\Gamma d$ with the quantum numbers of the Dirac matrix
$\Gamma$, the correlator is given by
\be
\label{mes_cor}
\Pi^{I=1}_\Gamma(\tau) &=& <{\rm Tr} [\Gamma S(0,\tau)\Gamma
  S(\tau,0)]> \, .
\ee
For isoscalar currents $j_\Gamma=\frac{1}{\sqrt{2}}(\bar u\Gamma u +
\bar d\Gamma d)$, there is an additional disconnected contribution
\be
\label{sing_cor}
\Pi^{I=0}_\Gamma(\tau) &=& <{\rm Tr} [\Gamma S(0,\tau)\Gamma
  S(\tau,0)]> - <{\rm Tr} [S(0,0)\Gamma ] {\rm Tr}
  [S(\tau,\tau)\Gamma]>\, .
\ee
Baryonic correlators are calculated in an analogous way from traces
involving three fermion propagators \cite{SSV_94}. The calculation of
the propagator in the multi-instanton background is described in
detail in \cite{SV_93a}. The most important point is that the zero
mode part of the Dirac operator is inverted exactly, implying that
the instanton induced ('t Hooft) interaction between quarks is
included to all orders. The averaging is performed over all
configurations in a given ensemble (random, quenched or unquenched).
This implies that while correlators in the quenched ensemble only
include diagrams in which the valence quarks (the quarks created
by the source) interact via the 't Hooft interaction, the unquenched
correlation functions also include vacuum bubbles to all orders
in the interaction. Although this will in general affect
all correlation functions, it is of particular importance in
channels like the $\eta'$ or the $\sigma$, where quarks can
annihilate into the vacuum.

   In the following, we will always normalize correlation functions
to the corresponding free correlator,
\be
\label{mes_cor_free}
\Pi_\Gamma^0(\tau) &=& {\rm Tr} [\Gamma S_0(\tau)\Gamma S_0(-\tau)] \,
\hspace{1cm}
S_0(\tau) = {\gamma_4\over 2\pi^2} \frac{1}{\tau^3} ,
\ee
If the ratio $R=\Pi_\Gamma(\tau)/\Pi_\Gamma^0(\tau) >1$, we
will refer to the correlator as attractive, while $R<1$ corresponds
to repulsive interactions.  The aim of this paper is not to make
large scale numerical calculations of masses and coupling constants of
hadrons, but rather to study the differences between the correlation
functions in the various ensembles. Nevertheless, in order to make
quantitative statements we need to determine these observables. In
practice, this is done using a simple ``zero width pole plus continuum"
model for the spectral functions. For a scalar meson, this corresponds
to the following parametrization of the coordinate space correlator
\be
\label{fit_cor}
\Pi_\Gamma(\tau)= \lambda^2_\Gamma D(m_\Gamma,\tau) +
  \int_{s_0}^\infty ds\, \rho_0(s)D(\sqrt{s},\tau) \, .
\ee
Here, $\rho_0(s)$ is the free spectral function corresponding to
the perturbative quark-antiquark bubble in the correlation function.
The resonance mass $m_\Gamma$, coupling constant $\lambda_\Gamma$
and continuum threshold $s_0$ are then extracted by fitting the
the parametrization (\ref{fit_cor}) to the measured correlation
function. Similar parametrizations for other mesonic and baryonic
correlation functions are given in \cite{SV_93b,SSV_94}, and a
brief summary of our conventions for the various currents and
coupling constants involved is given in table 2.

   In our previous work on the statistical mechanics of the instanton
liquid \cite{SS_95c} we have studied a number of different instanton
ensembles and the dependence of bulk parameters on the underlying
instanton interaction. Some of these parameters are summarized in
table 1. The first column shows the results obtained in fully
interacting simulations using the so called streamline interaction
(SLI). The second column shows our results using the same interaction
in the quenched approximation (for $N_f=0$). While the streamline
interaction has certain attractive features, it is not easily generalized
to finite temperature. For this reason we have also studied the
instanton ensemble using the ratio ansatz interaction (RAI). The
corresponding (unquenched) results are shown in the third column
while the last column shows the parameters of the random ensemble
for comparison. We should also emphasize that all of these ensembles,
with the exception of the random one whose parameters where determined
phenomenologically, satisfy a number of low energy theorems for
the topological susceptibility and fluctuations of the instanton
number, that follow from the renormalization properties of QCD
\cite{SS_95c}.

   All dimensionful quantities (again with the exception of the random
ensemble) are given in units of the (Pauli-Vilars) regulator $\Lambda$.
The value of $\Lambda$ has to be fixed from some physical observable
or by using the measured value of $\Lambda_{\overline{MS}}$. Since we
need a prescription to compare quenched and unquenched ensembles and
since the experimental accuracy of $\Lambda_{\overline{MS}}$ remains
poor, we have chosen to fix the physical instanton density at $n^{-1/4}
= 1\,{\rm fm}$. The corresponding scale parameters for the different
ensembles are listed in table 1. These values have been used to convert
the measured condensates into physical units, and will be used in the
following to provide the physical scale for the measured coordinate space
correlation functions.

   Before we come to a more detailed discussion of the correlation
functions we would like to discuss what differences between the
correlation functions in the different ensembles one would expect
on the basis of the global parameters listed in table 1. The random
ensemble, which is the simplest example of a quenched ensemble,
depends on only two parameters, the instanton density $n$ and the
diluteness $n\bar\rho^4$. Using the mean field approximation
\cite{Shu_82a,DP_86}, the dependence of the quark condensate,
the pion decay constant and the pion mass on these parameters
is given by
\be
\label{MFA}
\begin{array}{ccl}
  <\bar qq> & \sim & \sqrt{n}/\bar\rho , \\[0.2cm]
  f_\pi^2   & \sim & n\bar\rho^2
            \left(1+O\left(\log(\sqrt{n}\bar\rho^2)\right)\right),
            \\[0.2cm]
  m_\pi^2   & \sim & 1/(\sqrt{n}\bar\rho^3)
            \left(1+O\left(\log(\sqrt{n}\bar\rho^2)\right)\right).
\end{array}
\ee
If one uses the instanton density in order to fix the overall scale,
then we expect the pion decay constant to grow with the average
instanton size, while the quark condensate and the pion mass drop.
The quark constituent mass grows with the average size as $M\sim
\sqrt{n}\rho$. We have checked these predictions against numerical
calculations in the random ensemble and they agree fairly well.
These calculations also show that most hadronic masses, in particular
the rho meson, the nucleon and the delta do not scale like the constituent
mass but are essentially independent of $\bar\rho$. This provides further
evidence in favor of the hypothesis that these states are really
bound in the instanton model. An exception is the $a_1$ meson, whose
mass scales roughly like the constituent mass, and which is presumably
unbound in our model.

   We have calculated a large number of $I=1$ and $I=0$ mesonic correlators,
as well as those for nucleons and $\Delta$ baryons. Some of the results
are shown in fig.1 and the corresponding resonance parameters are summarized
in table 3. All of the correlators were obtained in a system of 128
instantons in a volume $(3.00)^3\times 4.74\,{\rm fm}^4$, with quark
masses $m_u=0.1 \Lambda$ and $m_s=0.7\Lambda$. In fig.1 the phenomenological
curves from \cite{Shu_93} are given by solid lines, and the results from the
random ensemble by the open triangles connected by dashed lines\footnote{The
results for the random ensemble differ slightly from the ones reported
in \cite{SV_93b,SSV_94}, mainly because they were obtained at a somewhat
larger quark mass in a smaller volume. The parameters used here were chosen
in order to facilitate comparison with the unquenched ensembles. For
unquenched calculations, the time and storage requirements for generating
the configurations limit us to smaller physical volumes.}. Our results for
the two interacting  ensembles are shown by the solid (SLI) and open (RAI)
data points.

    In fig.1a, we show the pion (pseudoscalar $I=1$) channel. Although
all ensembles show a qualitatively similar behaviour, the results in
the interacting ensembles lie systematically below the phenomenological
curve. There are two reasons for this deviation. The first one is just
a technical point: the light quark current masses in our simulations
are on the order of 20 MeV, still significantly heavier that the physical
value $(m_u+m_d)/2\simeq 6$ MeV. Using $m_\pi\sim \sqrt{m}$, as suggested
by the Gell-Mann, Oakes, Renner relation, one can extrapolate the pion
mass to the physical value of the quark masses. The results are also
given in table 3. We observe that the pion mass is smaller in the
interacting ensembles. The results are consistent with the experimental
value in the streamline ensembles (both quenched and unquenched), but
clearly too small in the ratio ansatz ensemble. This is in agreement
with the mean field result (\ref{MFA}), and a reflection of the fact
that the ratio ansatz ensemble is too dense.

   The other point is that the pion coupling to the pseudoscalar current,
$\lambda_\pi$, is related to the quark condensate and the pion decay
constant $\lambda_\pi=\langle\bar qq \rangle/f_\pi$. Since the quark
condensate in the unquenched ensembles is smaller as compared to the
random one, $\lambda_\pi$ is also reduced. This is a consequence of
correlations among instantons and not captured by the mean field
approximation.

   The pion decay constant $f_\pi$ can be determined in a number of
different ways, either from $\lambda_\pi$ using the measured quark
condensate and the PCAC relation quoted above, from the pion contribution
to the axial vector correlator, or from the off-diagonal pseudoscalar
axialvector correlator \cite{SV_93b}. All of these methods yield
consistent results and we simply quote an average in table 3. The
pion decay constant in the unquenched streamline ensemble is somewhat
too small, while in the ratio ansatz ensemble it is clearly too large.
The latter fact is again consistent with the mean field estimate
(\ref{MFA}) and related to the fact that the RAI ensemble is not
as dilute as phenomenology demands.

   In fig.1c we present the results for the isovector-vector and
axialvector ($\rho$ and $a_1$ meson) channels. These two correlation
functions are not affected by instantons to first order in the
instanton density. In this case the various ensembles lead to
results that are fairly similar to each other. The rho meson mass
in the unquenched ensemble is lighter than in the random ensemble
and now very close to the experimental value $m_\rho=770$ MeV.
Since the mass is essentially independent of the diluteness of
the ensemble, the difference must be due to correlations among
the instantons. How correlated instanton-antiinstanton pairs
may generate an attractive interaction in the vector channel
was discussed in \cite{SSV_95}. The $a_1$ mass in the interacting
ensembles, in particular ratio ansatz one, is too large as
compared to the experimental value. As mentioned above, the
$a_1$ is probably unbound in the instanton model and the
observed behavior is simply due to the constituent mass being
larger if the ensemble is very dense.

   The situation is drastically different for the two most repulsive
channels, the $\eta'$ meson (the $SU_f(3)$ singlet pseudoscalar) and
its chiral partner, the $I=1$ scalar which we will denote by $\delta$.
In Fig.1(b,d) we show the correlation functions for the non-strange
$\eta$ current $j_{\eta_{ns}}=\frac{1}{\sqrt{2}}(\bar u\gamma_5u+
\bar d\gamma_5 d)$ and the $\delta$ current  $j_\delta= \bar ud$. For
the delta, we do not have a phenomenological curve, but instead show
the lattice result \cite{CGHN_93b}.

   Out of about 40 correlation functions calculated in the
RILM \cite{SV_93b,SSV_94}, only the $\eta'$ and $\delta$ were
completely unacceptable (the dashed lines): these two correlation
functions drop very rapidly with distance, and become negative at
$x\sim 0.4 $fm. This behaviour is clearly incompatible with a normal
spectral representation of the correlators. The results in the
unquenched ensembles (closed and open points) significantly improve
the situation. The reason for this behavior is related to the formation
of instanton-antiinstanton pairs in the unquenched ensembles (or, more
generally, to the screening of the topological charge). For both the
$\delta$ and the $\eta'$ the single instanton contribution is repulsive,
but the contribution from pairs is attractive \cite{SSV_95}. Only if
correlations among instantons and antiinstantons are sufficiently strong,
the correlators are prevented from becoming negative. Quantitatively,
the $\delta$ and $\eta_{ns}$ masses in the streamline ensemble are still
too heavy as compared to their experimental values. In the ratio ansatz,
on the other hand, the correlation even show an enhancement at distances
on the order of 1 fm, and the fitted masses are too light. This shows
that these two channels are very sensitive to the strength of correlations
among instantons.

   In table 4 we summarize our results for baryons. The nucleon
parameters are not strongly affected by the choice of instanton ensemble.
The main difference is that the nucleon coupling constants are somewhat
smaller in the unquenched ensembles, an effect that is similar to
the reduction in $\lambda_\pi$ observed above. The behavior of the
Delta correlation functions is similar to the $\rho$ meson, with
the Delta mass being reduced in the unquenched ensembles. This
brings the measured Delta-nucleon mass splitting, which is too
large in the random ensemble, closer to its experimental value.

    To summarize the dependence of hadronic correlation functions
on the instanton ensemble, we find that pion properties are mostly
sensitive to global properties of the instanton ensemble, in particular
its diluteness. Good phenomenology demands $\bar\rho^4 n\simeq 0.03$,
as originally suggested in \cite{Shu_82a}. The properties of vector
mesons and the $\Delta$ are essentially independent of the diluteness,
but show some sensitivity to quark induced correlations. Correlations
induced by the fermion determinant are extremely important in heavy
scalar channels, like the $\eta'$ and $\delta$.

\section{Temperature dependence of hadronic correlation functions }

    In this section we would like to study the behavior of hadronic
resonances as the temperature is raised and the system undergoes the
chiral phase transition. The temperature dependence of mesonic
susceptibilities, defined as integrated meson correlation functions,
was already studied in \cite{SS_95c}. The main conclusion in that
work was that the transition is consistent with a nearby second
order phase transition for $N_f=2$. We could not conclusively
settle the question of the fate of the $U(1)_A$ symmetry above
the critical temperature of chiral symmetry restoration. Here,
we want to study the nature of the transition in much more detail,
by studying the behavior of hadronic correlation functions.

  Clearly, the behavior of hadronic correlation functions is
of great interest in connection with possible modifications of
hadrons in hot and dense matter \cite{AB_93,BR_91}. However,
there is very little phenomenological information about the
behavior of hadronic correlators at nonzero temperature. One
source of information is the Operator Product Expansion (OPE),
but at $T\neq 0$ additional assumptions about the temperature
dependence of the condensates are needed. Some consensus
has emerged from the work on sum rules based on the OPE
(see, for example \cite{BS_86,EI_95,HKL_93}). For most correlation
functions, the dominant power corrections are determined by
the quark condensate, and there is a tendency for resonance
masses and continuum thresholds to drop as chiral symmetry
is restored. Recently, the first results\footnote{There is
a large amount of work on screening masses at finite $T$, which
we will discuss in the next section.} from lattice measurements
of point-to-point correlation functions at $T\neq 0$ were
published in \cite{BGK_94}.

  At finite temperature, Lorentz invariance is broken, and the
correlation functions in the spatial and temporal direction
are independent. In addition to that, mesonic and baryonic
correlation functions have to obey periodic or antiperiodic
boundary conditions, respectively, in the temporal direction.
In the case of spacelike correlators one can still go to large
$x$ and filter out the lowest exponents known as screening masses.
While these states are of theoretical interest and have been studied
in a number of lattice calculations, they do not correspond to poles
of the spectral function in energy. In order to look for real bound
states, on has to study temporal correlation functions. However, at
finite temperature the periodic boundary conditions restrict the useful
range of temporal correlators to the interval $\tau<1/(2T)$ (about 0.6
fm at $T=T_c$), so that there is no direct procedure to extract
information about the groundstate. The underlying physical reason is
clear: at finite temperature excitations are always  present. In the
following we will study how much can be learned from temporal correlation
functions in the interacting instanton liquid. In the next section
we will also present the corresponding screening masses.

   The calculation of the correlation functions at non-zero temperature
is straightforward. As before, the correlators are determined from
ensemble averages of various contractions of the quark propagator.
The relevant instanton ensembles are the finite temperature ratio
ansatz ensembles discussed in detail in \cite{SS_95c}. The chiral
phase transition in this ensemble occurs at $T\simeq 125$ MeV. The
calculation of the quark propagator closely parallels the $T=0$ case.
The zero modes wave functions are substituted by their finite-$T$
counterparts \cite{Gro_77}, and the non-zero mode part is summed
over all fermionic Matsubara frequencies. As a result, the quark
propagator obeys antisymmetric boundary conditions in the finite
temperature box. Again, we normalize all correlation functions
to the corresponding noninteracting correlators, calculated from
the free propagator at finite temperature $T$
\be
\label{T_prop}
S_0(T,\tau) =  \frac{\gamma_4}{2\pi^2}\sum_{n=-\infty}^{\infty}
    \frac{(-)^n}{(\tau+n/T)^3} .
\ee
Resonance parameters at finite temperature are extracted from a
simple parametrization, analogous to the parametrization
(\ref{fit_cor}) discussed above. In the case of a scalar meson,
it reads
\be
\label{fit_Tcor}
\Pi_\Gamma(\tau) &=& \lambda^2_\Gamma D(T,m_\Gamma,\tau) +
{\rm Tr} [\Gamma S^V(T,m,\tau)\Gamma S^V(T,m,-\tau)] \, ,
\ee
where $D(T,m,\tau)$ is the finite temperature massive boson
propagator and $S^V(T,m,\tau)$ is the Dirac vector part of
the finite $T$ massive fermion propagator. The second term
corresponds to the contribution of massive constituent
quarks at finite temperature. As $m\to 0$, the continuum
threshold moves down to zero energy.

   The finite temperature temporal correlation functions in
the interacting instanton liquid are presented in figs.2-5. In
fig.2a,b we show the pion and sigma correlators for various
temperatures. Here and in the following figures, open points
correspond to temperatures below $T_c$, while the solid points
show results near and above $T_c$. The $\pi$ and $\sigma$
correlators are larger than the perturbative one at all
temperatures, implying that the interaction is attractive even
above $T_c$. The peak height decreases strongly with $T$,
but too a large part this is a simple consequence of the fact
that the length of the temporal direction shrinks. Below $T_c$,
the disconnected part of the sigma correlation functions tends
to the square of the quark condensate at large distance. Above
$T_c$, chiral symmetry is restored and the $\sigma$ and $\pi$
correlation functions become equal.

   The vector and axial vector correlation functions are shown
in fig.2(c,d). At low $T$ the two are very different while above
$T_c$ they become indistinguishable, again in accordance with
chiral symmetry restoration. In the vector channel, the changes
in the correlation function indicate the ``melting" of the resonance
contribution. At low $T$ (e.g. $T=0.43 T_c$, shown by the open
triangles) one can see a peak in the correlation function at $x
\simeq$1 fm, indicating the presence of a bound state well separated
from the two quark (or, more realistically, two pion continuum)
continuum. However, this signal disappears at $T\sim 100 MeV$,
implying that the $\rho$ meson coupling to local current becomes
small. This is consistent with the idea that the resonance ``swells''
(or completely dissolves) in hot and dense matter. A similar
phenomenon is also observed in the Delta baryon channel, see fig.3d.
At $T=0.43T_c$, there is a shoulder in the chiral even Delta
correlation function $\Pi_2^\Delta$, consistent with the
presence of a bound state, but the signal completely disappears
at $T=0.86 T_c$. Note however that in the instanton model, there
is no confinement and the amount of binding in the $\rho$ and
$\Delta$ channel is presumably small. In full QCD, these two
resonance might therefore be more stable as the temperature
increases\footnote{On the other hand, in the real world both
particles have a significant width, even at zero temperature.
The question whether they survive as (complex) poles will
therefore be difficult to answer.}.

     The dominant effect at small temperature is mixing between
the vector and axialvector channels \cite{DEI_90}. This means,
in particular, that there is a pion contribution to the vector
correlator at finite $T$. This contribution is most easily
observed in the longitudinal vector channel $\Pi^V_{44}(\tau)$
(note that in fig.2 we show the trace $\Pi^V_{\mu\mu}(\tau)$
of the vector correlator). We find a sizeable enhancement in
this channel at $T=77$ MeV, but the effect disappears at larger
temperatures $T>110$ MeV.

     For $T>110$ MeV both the $\rho$ and $\Delta$ correlation functions
are well described by a continuum contribution only, modeled by the
propagation of two or three massive quarks. The temperature dependence
of the effective quark mass extracted from fits to the correlators is
shown in fig.4. We observe that the values determined from the $\rho$
meson and $\Delta$ baryon channels are in very good agreement. Note that
the effective quark mass does not vanish at $T=T_c$. This behavior does
not contradict chiral symmetry restoration. The effective mass consist
of two parts, a scalar component which does violate chiral symmetry
and indeed disappears at the critical point, and a vector component
which is not constrained by the symmetry. The scalar part is mostly
due to spontaneous symmetry breaking in the random instanton liquid,
while the vector component also receives contributions from nonzero
modes in the molecular vacuum.

      At this point, let us compare our results to those obtained on
the lattice. Point-to-point correlation functions at $T\neq 0$ in the
temporal direction were calculated by Boyd et al.~\cite{BGK_94}. The
lattice used in this work has $N_\tau=8$ points in time direction
(and $N_\sigma=16$), so the information is clearly very limited.
Still, one observes that the vector correlation function behaves
perturbatively (as two independently propagating quarks) while the
pion one does not, even above $T_c$. If we normalize the correlators
so that $\Pi_\pi(\tau)/\Pi_\rho(\tau)=1$ at $\tau=0$, then we find
that at $\tau=0.5T^{-1}$ (where the effect is maximal) and $\beta=5.3$
(corresponding to $T\simeq 1.2 T_c$) this ratio is $\Pi_\pi(\tau)/
\Pi_\rho(\tau) \simeq(2-4)$. This can be compared to our results shown
in fig.2, where at $T=1.13 T_c$ this ratio is about 1.5. So, the
agreement is quite fair, taking into account that the lattice data
were obtained with only 8 points in the temporal direction\footnote{
And these 8 points, according to our model, represent gauge field
configurations corresponding to  instanton-antiinstanton molecules
polarized in the temporal direction!}.

    In figure 3, we show our results for some of the baryon correlation
functions. As explained in detail in \cite{SV_93b}, and summarized briefly
in table 5, there are 6 nucleon and 4 delta correlation functions
that can be constructed out of the two Ioffe currents for the nucleon
and the unique delta current. Half of these correlators, $\Pi_{1,3,5}^N$
and $\Pi_{1,3}^\Delta$ are chiral odd and have to vanish as chiral symmetry
is restored. $\Pi_{2,4}^N$ as well as $\Pi_{2,4}^\Delta$ are chiral even
and may show resonance signals even as chiral symmetry is restored.
$\Pi_6^N$, the chiral even, off-diagonal correlator of the two Ioffe
currents is special. Chiral symmetry does not require it to vanish, but
since the two currents have different $U(1)_A$ transformation properties,
it is sensitive to $U(1)_A$ symmetry breaking \cite{SS_95b}.

   The chiral-odd correlator $\Pi_1^N$ is shown in fig.3a. Since it does
not receive a contribution from free quark propagation, the correlation
function (in our normalization) starts at 0 at $\tau=0$. At small
temperature, there is a large signal which is dominated by the nucleon
contribution even at fairly small distances $\tau\simeq 0.3$ fm. In
accordance with chiral symmetry restoration, this signal disappears
as $T\rightarrow T_c$. In terms of the nucleon spectrum, there are
several possible explanations for this behavior. If one includes the
lowest positive ($N^+$) and negative parity ($N^-$) resonances in the
spectral function, we have
\be
\Pi_1^N(\tau) = (\lambda_1^{N^+})^2 m_{N^+}D(m_{N^+},\tau)
               -(\lambda_1^{N^-})^2 m_{N^-}D(m_{N^-},\tau),
\ee
where $\lambda_1^{N^\pm}$ is the coupling of the first Ioffe current
to the positive/negative parity nucleon. This implies that the correlator
can vanish either because (i) the masses go to zero $m_{N^+}=m_{N^-}=
\ldots=0$, or (ii) the resonances decouple $\lambda_1^{N^+}=\lambda_1^
{N^-}=\ldots=0$, or (iii) the parity partners become degenerate $m_{N^+}
=m_{N^-},\;\lambda_1^{N^+}=\lambda_1^{N^-}$ and their contributions to
the correlator cancel each other.

     In order to distinguish between these possibilities let us consider
the chiral-even nucleon correlator $\Pi_2^{N}$ (see fig.4b). There is an
enhancement in $\Pi_2^N$, which is fairly stable as a function of
temperature. Since both parity states contribute with the same sign
to $\Pi_2^N$ (see the summary in table 5), the possibility (ii) mentioned
above does not seem likely. The corelation function $\Pi_6^N$ shown in
fig.4(c) is very large at low temperature, but strongly suppressed near
and above the chiral phase transition. Similar to other correlation
functions sensitive to the fate of the $U(1)_A$ symmetry, which we will
discuss below, this phenomenon depends strongly on the value of the
current quark masses. In order for $\Pi_6^N$ to vanish, the contribution
from different resonances have to cancel each other, suggesting the
existence of degenerate states with different parity.

   Clearly, the numerical values of the nucleon mass and its coupling
constant are very interesting. Unfortunately, our ignorance concerning
the form of the spectrum and the limited information provided by the
temporal correlation functions makes it very difficult to provide a
definite result. The discussion above clearly suggests that we should
a least include two (opposite parity) resonances in addition to the
continuum contribution. Furthermore, the masses seen in the
scalar and vector channel are independent of each other. As a result,
there are too many parameters in order to sufficiently constrain any
fit. The correlation functions in the chiral odd channels $\Pi_{1,3,5}^N$
can be described both in terms of a dropping scalar mass, or using
two (almost) degenerate states with different parity. The chiral even
correlators $\Pi_{2,4}^N$ are somewhat more restrictive. We find
that the correlation functions require a nucleon mass (or vector
self energy) and coupling that is almost independent of temperature
while the continuum threshold drops.

    Let us finally discuss the fate of the $U(1)_A$ symmetry from the
correlation functions in the mesonic sector. Here we consider the $\delta$
and $\eta_{ns}$ channels, the $U(1)_A$ partners of $\pi$ and $\sigma$.
The temperature dependence of these correlation functions is shown in
fig.5 (a,b). We remind the reader that the amount of repulsion in the
$\delta$ and $\eta_{ns}$ was too strong in the random model (dashed
lines), but the behavior is improved in the interacting model. In
contrast to the cases considered above in which correlation functions
at finite temperature were reduced in magnitude, now one observes
the opposite trend. As the temperature increases, the amount of
repulsion is clearly reduced. This effect is more pronounced in
the $\delta$ than in the $\eta_{ns}$ channel, partly because the
$T=0$ ensemble is overcorrelated and produces an $\eta_{ns}$ mass
that is already too light.

  The question of $U(1)_A$ restoration at $T=T_c$ (or above) can be
studied by comparing the $\eta$ and $\delta$ correlators with
their $U(1)_A$ partners, the $\pi$ and $\sigma$ (shown in fig.2).
A serious complication for this comparison is the strong dependence
of the $\eta$ and $\delta$ correlators on the current quark mass. In
fig.5 (d) we show the $\delta$ correlator at fixed temperature $T=0.86
T_c$ for a number of different current quark masses\footnote{These
numbers refer to the valence quark mass, used in the calculation
of the fermion propagator. The sea quark mass, used in generating
the instanton ensemble, was fixed at 13 MeV.} $m=5.2, 8.6, 13, 26$
MeV. For a very small quark mass $m=5.2$ MeV (open triangles), there
is a significant enhancement in the $\delta$ channel, which is indeed
equal to that in the pion channel. This would imply chiral and $U(1)_A$
restoration. However, this is not a physical result, but a finite
size effect. If the quark masses in a system with finite volume
become too small, chiral (and $U(1)_A$) symmetry breaking are lost.

   In order to be more quantitative one would like to determine
the precise masses and coupling constants of the $\eta_{ns}$ and
$\delta$. Again, given the restricted information available and
the uncertainty concerning the form of the spectrum, it is hard
to obtain reliable results for these parameters. In practice, we
need to make some assumptions. Here, we first fix the non-resonant
continuum, assuming it to be given by free propagation of quarks
with the effective masses determined above. In the temperature
region of interest, $T\geq 100$ MeV, the pion and delta masses
are roughly consistent with zero $m_\pi\simeq m_\delta\simeq 0\pm
200$ MeV. In the following we therefore fix the masses to
be zero and study the temperature dependence of the coupling constants
$\lambda_{\pi,\,\delta}$. In fig.5(c) we compare the coupling
constants for the pion (stars) and delta. At $T=130$ MeV we show
the value of $\lambda_\delta$ for the different quark masses
discussed above. For the lowest mass $m=5.2$ MeV, the $U(1)_A$
symmetry is restored even below $T_c$, but this is definitely
a finite size effect. For a somewhat larger mass $m=8.6$ MeV,
$U(1)_A$ symmetry appears to be restored at or just above $T_c$.
At even larger current quark masses, no evidence for $U(1)_A$
restoration is seen. To settle this important question will
require significantly larger simulations.

   Possible experimental signatures of (partial) $U(1)_A$ chiral
symmetry restoration were discussed in \cite{Shu_94} and, more recently,
in \cite{KKL_95,HW_95}. Basically, the idea is that if the $\eta_{ns}$
becomes degenerate with the $\pi$ at $T\simeq T_c$, one would expect
a significant (low $p_T$) enhancement of $\eta$ and $\eta'$ production
in relativistic heavy ion collisions. The more recent papers have
pointed out two important experimental hints. First, as discussed in
\cite{HW_95}, the WA80 collaboration at CERN found a (modest)
enhancement of the $\eta/\pi^0$ ratio in central relative to
peripheral (or pp) collisions \cite{WA80_94}. Furthermore, the
enhancement of the low-mass dilepton production observed by the
CERES \cite{CER_95c} and Helios \cite{HEL_95} collaborations could be
explained in terms of the Dalitz decays of (more abundant) $\eta'$
mesons \cite{KKL_95}. To study these suggestions more quantitatively
would not only require a better determination of the $\eta_{ns}$ mass,
but also an understanding of $\eta-\eta'$ mixing and $\eta'$ absorption
as a function of temperature.

    Summarizing this section we find that many correlation
functions\footnote{At least those correlation functions, that
are not required by chiral symmetry to vanish at the phase
transition.}
(in the restricted range in which they are accessible at finite
temperature) are remarkably smooth as a function of temperature,
despite the fact that the  vacuum fields and the quark condensate
change significantly. Phenomenologically this means that
the melting of resonances
is compensated by the continuum threshold moving down in energy.
Roughly speaking, we find three different types of behavior in
chiral even correlators. In ``attractive" channels, like the $\pi$ and
$\sigma$ meson or the nucleon, resonance contributions seem to survive
the phase transition. In channels that do not have strong interactions
at $T=0$ (like the $\rho$ meson and the $\Delta$), the resonances
disappear quickly, and the correlators can be described in terms
of free quark propagation with a certain effective chiral mass.
Finally, the most dramatic change is seen in ``repulsive" channels
$\delta,\eta'$, where most of the repulsive interaction disappears
near the chiral phase transition.

\section{Screening masses}

   Correlation functions in the spatial direction can be studied
at arbitrarily large distance, even at finite temperature. This
means that contrary to the temporal correlators, the corresponding
spectrum can be determined with good accuracy. Although this
spectrum is not directly related to the spectrum of physical
excitations,  the structure of spacelike screening
masses is still of theoretical interest and has been investigated
in a number of lattice \cite{TK_87,Goc_91} and theoretical
\cite{EI_88,HZ_92,KSB_92,HSZ_94} works.

   At finite temperature, antiperiodic boundary conditions in the
temporal direction require the lowest Matsubara frequency for fermions
to be $\pi T$. This energy acts like a mass term for propagation in
spatial direction, so quarks effectively become massive. At asymptotically
large temperatures, quarks only propagate in the lowest Matsubara mode,
and the theory undergoes dimensional reduction \cite{AP_81}. The spectrum
of spacelike screening states is then determined by a 3-dimensional
theory of quarks with chiral mass $\pi T$, interacting via the
3-dimensional Coulomb law and the nonvanishing spacelike string
tension \cite{Bor_85,MP_87}.

   Dimensional reduction at large $T$ predicts almost degenerate
multiplets of mesons and baryons with screening masses close to
$2\pi T$ and $3\pi T$. The splittings of mesons and baryons with
different spin can be understood in terms of the nonrelativistically
reduced spin-spin interaction. The resulting pattern of screening
states is in qualitative agreement with lattice results even at
moderate temperatures $T\simeq 1.5 T_c$. The most notable exception
is the pion, whose screening mass is significantly below $2\pi T$.

   In this section we would like to study whether the spectrum of
screening states can also be understood in the instanton liquid model.
One should note that dimensional reduction does not naturally occur in
the instanton model. Instantons have fermionic zero modes at arbitrarily
large temperature, and the perturbative Coulomb and spin-spin interactions
mentioned above are not included in our model.

   The calculation of the spatial correlation functions is performed
in the same way as described for the temporal correlators in the last
section. Technically, there is no differences between the two cases,
except that one has to make sure that there is no sensitivity
to boundary effects in the spatial direction, while the boundary
conditions in the temporal condition are of course essential.

   We have not made an attempt to provide multi-parameter fits to
the spatial correlation functions. All we are interested in are the
corresponding screening masses $\Pi(r)\stackrel{r \to \infty}\longrightarrow
\exp(-m_{scr}r)$. Our results are summarized in fig.6, where we have
already normalized the data to the lowest fermionic Matsubara frequency
$\pi T$. We find a picture that is qualitatively quite similar to
our earlier work in the context of a schematic ``cocktail" model
(see fig. 11 of \cite{SSV_95}), in which the instanton ensemble near
$T_c$ is described as a mixture of a random and a molecular instanton
liquid.

First of all, the screening masses clearly show the restoration
of chiral symmetry as $T\to T_c$: chiral partners like the $\pi$ and
$\sigma$ or the $\rho$ and $a_1$ become degenerate. Furthermore,
the mesonic screening masses are close to $2\pi T$ above $T_c$,
while the baryonic ones are fairly close to $3\pi T$, as expected.
Most of the screening masses are shifted slightly upwards as
compared to the most naive prediction. Considering the vector
channels $\rho,a_1,\Delta$, this shift corresponds to a residual
``chiral quark mass" on the order of 120-140 MeV.

    The most striking observation is the strong deviation from this
pattern seen in the scalar channels $\pi$ and $\sigma$, with screening
masses significantly below $2\pi T$ near the chiral phase transition.
This effect persists to fairly large temperature. We also find that
the nucleon-delta splitting does not disappear at the phase transition,
but decreases smoothly. We have also studied different components of
the vector correlator. The results shown by the full points in fig.6
have been extracted from the trace of the vector correlator,
$\Pi_{\mu\mu}\sim \exp(-m_\rho r)$. Alternatively, one can study the
longitudinal $\Pi_{44}\sim\exp(-m_{\rho_4}r)$ or transverse $\Pi_{ii}
\sim \exp(-m_{\rho_i}r)$ components of the vector correlator. The
results are also shown in fig.6. We find that the longitudinal
screening mass near the chiral phase transition is smaller than the
transverse one, $m_{\rho_i}-m_{\rho_4}\simeq 100$ MeV. This result is
in qualitative agreement with the prediction from dimensional reduction
\cite{HSZ_94}, but disagrees with the lattice calculation \cite{TK_87}.
Note that in the limit of a dilute system of fully polarized molecules,
$\rho_4$ is expected to be degenerate with the pion \cite{SSV_95}. The
presence of unpaired instantons, interactions among the molecules and
deviations from complete polarization cause the pion to be significantly
lighter than the longitudinal rho.

\section{The chirally symmetric phase in multi-flavor QCD}

    QCD like theories with different matter content form a rich
field and an interesting testing ground for our understanding of
nonperturbative phenomena in QCD. Significant progress in this
direction has recently been made in supersymmetric extensions of
QCD \cite{Sei_94}, where it has been possible to determine the
critical number of flavors, such that chiral symmetry is restored
in the ground state for $N_f>N_f^{crit}$. Furthermore, many
interesting phenomena, like exact nonabelian dualities occur
in theories with specific flavor content. In our previous work
\cite{SS_95c} we have studied the phase diagram of QCD with many
flavors (in the instanton model). We found that chiral symmetry
is restored in the groundstate for more than four light quark
flavors (in the case $N_c=3$). In this section we would like to
study the properties of this new phase in more detail, by
considering hadronic correlation functions.

    One more reason to look into spectrum of multi-flavor QCD
with a chirally symmetric groundstate is the fact that it might
provide some insight into the spectrum of QCD above the phase
transition. In the instanton model, the mechanism for the large
$N_f$ transition is very similar to the one that causes the phase
transition at finite temperature, the formation of correlated
instanton-antiinstanton pairs\footnote{The main difference is
that now we consider T=0 case, so Lorentz invariance is not broken and
there is no preferred direction for the formation of molecules.}.
Although correlation functions in the temporal direction at finite
$T$ shed some light on the nature of hadronic excitations at large
temperature, it is hard to extract definite results. In the chirally
symmetric phase for large $N_f$, on the other hand, one can study
correlation functions at large distances and the determination of
the spectrum is comparatively easy.

    In \cite{SS_95c} we found that the critical number of massless
flavors in the instanton model is close to $N_f=4$. Here we will
consider the case $N_f=5$, in order to be sure to be in the chirally
restored phase. In this case the instanton density is $n=0.0097
\Lambda^4$, significantly smaller than the $N_f=2$ case studied
above: adding more light quarks reduces the tunneling probability.
In the following, we will study correlation functions as a function
of $x$ in units\footnote{In principle one could fix units as in section
2 above, but since $N_f=5$ QCD is not a physical theory, this would not
make much sense.} of $\Lambda^{-1}$. Since $n$ in units of $\Lambda^4$
is so much smaller than before, the typical distance will be rather
large.

    The correlation functions are calculated as described in section 2.
Some of the results are shown in fig.6. Despite the fact that the distance
$x$ is unusually large, the deviation of the correlators from free
propagation of massless quarks is only at the 10\% level. The results for
the vector and axial vector currents (which we label, as in QCD, by
$\rho$ and $a_1$) are quite typical for most correlators, including
baryons. The correlators are suppressed with respect to free propagation,
and their functional form can be explained simply in terms of a nonzero
quark mass. Furthermore, this quark mass is consistent with the input
quark mass, there is no dynamical mass generation. The splitting between
the $\rho$ and $a_1$ correlators is also consistent with the symmetry
breaking caused by the nonzero current masses.

   The only channels that display a non-trivial behavior are the scalar
and pseudoscalar mesons. In fig.6 we show the $\pi$ and $\delta$ correlation
functions, both of which show an attractive interaction. If we interpret
this enhancement as in eq. (\ref{fit_Tcor}) in terms of a resonance
superimposed on a two-quark continuum, we find
\be
m_\pi=(1.4\pm 0.38)\Lambda,\hspace{0.5cm}
\lambda_\pi=(0.24\pm 0.05)\Lambda^2, \hspace{0.5cm}
m_{q}=(0.10 \pm 0.1)\Lambda \\
m_\delta=(1.4\pm 0.42)\Lambda, \hspace{0.5cm}
\lambda_\delta=(0.22\pm 0.05)\Lambda^2, \hspace{0.5cm}
m_{q}=(0.13 \pm 0.1)\Lambda .
\ee
Again, the quark masses are consistent with the input current
masses. Furthermore, the $\pi$ and $\delta$ mesons appear to be
degenerate. This is not unexpected in the multi-flavor theory,
since in the absence of chiral symmetry breaking the $2N_f$-leg
't Hooft vertex cannot directly contribute to a mesonic two
point function. Thus, we conclude that the spectrum in this
phase consists of a multiplet of weakly bound scalar-pseudoscalar
resonances coupled to (almost) massless quarks. These states are
of course not Goldstone modes, but massive states, bound by the
molecule induced interaction. They are also not stable particles,
but resonances that decay into quarks.

\section{Conclusions and discussion}

    In summary, we have studied hadronic correlation functions in
the interacting instanton liquid. At zero temperature, we have
considered the behavior of hadronic correlators in different
instanton ensembles, in particular the influence of quenching
on the hadronic spectrum. While there are some subtle differences
in the pion and rho meson parameters\footnote{We should note,
however, that the scale parameters in the quenched and unquenched
ensembles differ significantly. The more accurate statement is
therefore that most correlation functions in the quenched and
unquenched ensembles only differ by a common change of scale.},
the main effect of the fermion determinant is seen in the $\eta'$
and $\delta$ mesons channels. These are the only channels in which
the random model fails completely.

     Correlations between instantons
induced by the fermionic determinant weaken the strong repulsion
observed in the quenched (and random) ensembles and clearly
improve the description of these two channels. However, a
quantitative description of the $\eta'$ and $\delta$ masses
depends on details of the instanton interaction that are not
well determined (and understood). Correlations between instantons
are somewhat too weak (and the $\eta'$ too heavy) in the streamline
ensemble, stabilized by the phenomenological core determined
in our previous work, while the correlations are too strong
(and the $\eta'$ is too light) in the ratio ansatz ensemble.
An important global parameter of the instanton liquid is its diluteness.
Good phenomenology, in particular in the pion channel, requires
the instanton liquid to be rather dilute $\bar\rho^4n\simeq 0.03$.
The streamline ensembles, both quenched and unquenched, approximately
satisfy this requirement, while the ratio ansatz leads to an ensemble
which is  too dense.

    We have also studied temporal and spatial correlation
functions in the instanton vacuum at finite temperature. Both
types of correlators clearly show the restoration of chiral
symmetry at $T\simeq 125$ MeV. In the instanton liquid model,
the mechanism for this transition is the formation of
instanton-antiinstanton molecules. This implies that even
above the chiral phase transition there is a significant
density of instantons, and we can expect nonperturbative
contributions to the correlation functions. We indeed find such
effects, most notably in the temporal correlation functions of
light scalar mesons. In channels like the $\pi$ and $\sigma$ we
find stable resonance contributions that appear to survive
the chiral phase transition. This result is in agreement with
a recent lattice calculations of temporal correlators \cite{BGK_94},
where a similar enhancement in the scalar-pseudoscalar mesons
channels in the chirally symmetric phase was observed.

     Contrary to the light scalar mesons (and the nucleon channel),
resonance contributions in the vector meson (and delta baryon)
channels decrease very fast as the temperature increases. Above
$T\simeq 100$ MeV, the correlation functions are consistent with
free propagation of quarks with a nonvanishing chiral mass. This
mass varies smoothly through the phase transition, and only
disappears at significantly larger temperatures. Finally, in
the heavy scalar meson channels, in particular the $\eta'$ and
$\delta$, the strong repulsion seen at $T=0$ disappears. Near
the chiral phase transition, we observe some tendency toward
$U(1)_A$ symmetry restoration. This result, however, depends
very sensitively on the value of the current quark mass. A final
verdict on the fate of $U(1)_A$ symmetry will therefore require
detailed simulations in much larger volumes.

   As already emphasized in the introduction, our study of
correlation functions at finite temperature is qualitative
in nature. We have been unable to give detailed numerical
predictions of the temperature dependence of hadronic masses
and coupling constants. This problem is related to the limited
range covered by temporal correlation functions, our ignorance
concerning the form of the spectral density, and the fact that
more resonances need to be included at $T\neq 0$. It is a difficulty
that our approach shares with other attempts, like QCD sum rules
or the lattice, to determine the hadronic spectrum at finite
temperature. Only more accurate results, the calculation of as
many independent correlators as possible, and more theoretical
work on possible modes of chiral restoration in the hadronic
spectrum will resolve this important question.

  An example for the necessity to consider more than just one
resonance is provided by the nucleon channel. We have considered
six different nucleon correlation functions with different chiral
symmetry properties. We conclude that, most likely, parity doubling
plays a role in restoring the symmetry in the nucleon spectrum. The
vector self energy stays roughly constant as a function of
temperature, while it is difficult to observe a dropping scalar
self energy.

     In order to provide a more detailed comparison of our model
with existing lattice calculations, we have also determined the
spectrum of spacelike screening masses. The pattern of screening
masses is much easier to determine than the spectral function in
energy, and most mesonic and baryonic states at $T>T_c$ are close
to simple multiples of the lowest Matsubara frequency $\pi T$.
The most important exception are the light scalars $\pi$ and $\sigma$,
whose screening mass at $T_c$ is only about half the naive value
of $2\pi T$. We find a nonzero nucleon-delta splitting, as well
as a splitting between the longitudinal and transverse components
of the vector mesons. The latter appears to be reversed as compared
to lattice simulations.

    Finally, we have studied hadronic correlation function in QCD
with many light flavors. This is a rich new field, that may provide
many interesting lessons. Above a certain critical number of light
flavors ($N_f^{crit}=4$ in our case), chiral symmetry is restored
even in the groundstate. In this case it is easier to determine
the spectrum, since one can follow correlation functions to large
distance. We find evidence of a (non Goldstone) ``pion" resonance
in multiflavor QCD, which is degenerate with the scalar sigma
and delta mesons. Most likely, these mesons are organized in
$U(N_f)\times U(N_f)$ multiplets. This strange world may offer
some clues about the spectrum of ordinary QCD above the phase
transition\footnote{However, there are also important differences.
For example, in multiflavor QCD the 't Hooft vertex has ``too many
legs", so that it cannot contribute directly to meson correlation
functions. We may therefore have a situation in which $U(1)_A$
symmetry is broken, but the $\eta'$ is degenerate with the pion.}.

\section{Acknowledgements}
We would like to thank J. Verbaarschot for many useful discussions.
This work was supported in part by US DOE grant DE-FG-88ER40388 and
DE-FG06-90ER40561. Some of the numerical calculations were performed
at NERSC, Lawrence Livermore Laboratory.

\newpage



\newpage\noindent
\begin{figure}
\caption{
Mesonic point-to-point correlation functions as a function of distance. All
correlators are normalized to free quark propagation. Fig.1(a) shows the
pseudoscalar (pion) correlator, (b) the isoscalar $\eta_{ns}$, (c) the vector
and axialvector correlators ($\rho,a_1$) and (d) the isovector scalar
($\delta$). The correlators are shown in different instanton ensembles,
the unquenched streamline ensemble (solid points), the ratio ansatz ensemble
(open points), and the random ensemble (dashed lines). The solid lines
correspond to the phenomenological results, except for fig.1(d) which
shows the result of a quenched lattice calculation.}
\end{figure}

\begin{figure}
\caption{Temporal correlation functions at $T\neq 0$, normalized to free
thermal correlators. Fig.2(a) shows the pseudoscalar (pion) correlator,
(b) the isoscalar scalar $\sigma$, (c) the isovector axialvector ($a_1$)
and (d) the isovector vector ($\rho$).  Correlators in the chirally symmetric
phase ($T \geq T_c$) are shown as solid points, below $T_c$ as open points.
The open triangles, squares and hexagons correspond to $T=0.43$, 0.60 and
0.86 $T_c$, while the closed triangles and squares show the data at $T=1.00
\,T_c$ and 1.13 $T_c$. For comparison we show the phenomenological $T=0$
results from fig.1 (solid lines).}
\end{figure}

\begin{figure}
\caption{Baryon temporal correlation functions at non-zero temperature.
The curves are labeled as in Fig.2. Figures (a), (b) and (c)show the
nucleon correlators $\Pi_1^N, \Pi_2^N$ and $\Pi_6^N$. Figure (d) shows
the chiral even Delta correlator $\Pi_2^\Delta $.}
\end{figure}

\begin{figure}
\caption{Temporal correlation functions at $T\neq 0$ for the nonstrange
$\eta$ (a) and the isovector scalar $\delta$ current (b). The curves are
labeled as in fig.2,3. Figure (d) shows the scalar ($\delta$) correlator
at fixed temperature $T=0.86 T_c$ for different quark masses, $m=5.2,8.6,
13, 26$ MeV, shown by open squares and triangles, closed squares and hexagons,
respectively. In fig.(c) we show the value of the pseudoscalar and scalar
coupling constants as a function of temperature. At $T=0.86 T_c$ we show
several values of $\lambda_\delta\, [{\rm fm}^{-2}]$, corresponding to the
different quark masses used in fig.(d).}
\end{figure}

\begin{figure}
\caption{Effective ``chiral" quark mass as a function of temperature $T$.
The open (closed) points correspond to the values extracted from the
$\Delta$ baryon and $\rho$ meson correlators.}
\end{figure}

\begin{figure}
\caption{Spectrum of spacelike screening masses as a function of
temperature. The masses are given in units of the lowest fermionic
Matsubara frequency $\pi T$.}
\end{figure}

\begin{figure}
\caption{Correlation functions at zero temperature in $N_f=5$ QCD.
The correlators are normalized to free propagation and the distance
is given in units of the scale parameter.}
\end{figure}

\setcounter{figure}{0}
\newpage

\begin{table}
\begin{tabular}{crrrr}
               & streamline         & quenched           & ratio ansatz
               & RILM              \\  \tableline
$n$            & 0.174$\Lambda^4$   & 0.303$\Lambda^4$   & 0.659$\Lambda^4$
               & 1.0  ${\rm fm}^4$ \\
$\bar\rho$     & 0.64$\Lambda^{-1}$ & 0.58$\Lambda^{-1}$ & 0.66$\Lambda^{-1}$
               & 0.33 ${\rm fm}$   \\
               & (0.42 fm)          & (0.43 fm)          & (0.59 fm)
               &                   \\
$\bar\rho^4 n$ & 0.029              & 0.034              & 0.125
               & 0.012             \\
$<\bar q q>$   & 0.359$\Lambda^3$   & 0.825$\Lambda^3$   & 0.882$\Lambda^3$
               & $(264\,{\rm MeV})^3$ \\
               &$(219\,{\rm MeV})^3$&$(253\,{\rm MeV})^3$&$(213\,{\rm MeV})^3$
               &             \\
$\Lambda$      & 306 MeV            &    270 MeV         &   222 MeV
               &  -          \\
\end{tabular}
\caption{Bulk parameters of the different instanton ensembles.}
\end{table}

\begin{table}
\begin{tabular}{cccl}
channel         &    current
                & matrix element
                & experimental value    \\  \tableline
$\pi$           &  $j^a_\pi=\bar q\gamma_5\tau^a q$
                &  $<0|j_\pi^a|\pi^b>=\delta^{ab}\lambda_\pi$
                &  $\lambda_\pi\simeq (480\,{\rm MeV})^2$     \\
                &  $j^a_{\mu\,5}=\bar q\gamma_\mu\gamma_5\frac{\tau^a}{2}q$
                &  $<0|j^a_{\mu\,5}|\pi^b>=\delta^{ab}q_\mu f_\pi$
                &  $f_\pi=93$ MeV \\
$\delta$        &  $j^a_\delta=\bar q\tau^a q$
                &  $<0|j_\delta^a|\delta^b>=\delta^{ab}\lambda_\delta$
                &       \\
$\sigma$        &  $j_\sigma=\bar q q$
                &  $<0|j_\sigma|\sigma>=\lambda_\sigma$
                &       \\
$\eta_{ns}$     &  $j_{\eta_{ns}}=\bar q\gamma_5  q$
                &  $<0|j_{\eta_{ns}}|\eta_{ns}>=\lambda_{\eta_{ns}}$
                &       \\
$\rho$          &  $j^a_{\mu}=\bar q\gamma_\mu\frac{\tau^a}{2}q$
                &  $<0|j^a_{\mu}|\rho^b>=\delta^{ab}\epsilon_\mu
                   \frac{m_\rho^2}{g_\rho}$
                &  $g_\rho=5.3$  \\
$a_1$           &  $j^a_{\mu\,5}=\bar q\gamma_\mu\gamma_5\frac{\tau^a}{2}q$
                &  $<0|j^a_{\mu\,5}|a_1^b>=\delta^{ab}\epsilon_\mu
                   \frac{m_{a_1}^2}{g_{a_1}}$
                &  $g_{a_1}=9.1$  \\
$N$             &  $\eta_1 = \epsilon^{abc}(u^aC\gamma_\mu u^b)\gamma_5
                   \gamma_\mu d^c$
                &  $<0|\eta_1 |N(p,s)> =\lambda_1^N u(p,s)$
                &             \\
$N$             &  $\eta_2 = \epsilon^{abc}(u^aC\sigma_{\mu\nu} u^b)
                   \gamma_5\sigma_{\mu\nu} d^c$
                &  $<0|\eta_2 |N(p,s)> =\lambda_2^N u(p,s)$
                &             \\
$\Delta$        &  $\eta_\mu = \epsilon^{abc}(u^aC\gamma_\mu u^b) u^c$
                &  $<0|\eta_\mu |N(p,s)> =\lambda^\Delta u_\mu(p,s)$
                &             \\
\end{tabular}
\caption{Definition of various currents and matrix elements used
in this work.}
\end{table}

\begin{table}
\begin{tabular}{ccrrrr}
               &                    & streamline         & quenched
               & ratio ansatz       & RILM   \\  \tableline
$m_\pi$        & $[\rm GeV]$        &  0.265             &   0.268
               &  0.128             &  0.284      \\
$m_\pi$ (extr.)& $[\rm GeV]$        &  0.117             &   0.126
               &  0.067             &  0.155      \\
$\lambda_\pi$  & $[{\rm GeV}^2]$    &  0.214             &   0.268
               &  0.156             &  0.369      \\
$f_\pi$        & $[\rm GeV]$        &  0.071             &   0.091
               &  0.183             &  0.091      \\
$m_\rho$       & $[\rm GeV]$        &  0.795             &   0.951
               &  0.654             &  1.000      \\
$g_\rho$       &                    &  6.491             &   6.006
               &  5.827             &  6.130      \\
$m_{a_1}$      & $[\rm GeV]$        &  1.265             &   1.479
               &  1.624             &  1.353      \\
$g_{a_1}$      &                    &  7.582             &   6.908
               &  6.668             &  7.816      \\
$m_{\sigma}$   & $[\rm GeV]$        &  0.579             &   0.631
               &  0.450             &  0.865      \\
$m_{\delta}$   & $[\rm GeV]$        &  2.049             &   3.353
               &  1.110             &  4.032      \\
$m_{\eta_{ns}}$& $[\rm GeV]$        &  1.570             &   3.195
               &  0.520             &  3.683 \\
\end{tabular}
\caption{Meson parameters in the different instanton ensembles. All
quantities are given in units of GeV. The current quark mass is $m_u
=m_d=0.1\Lambda$. Except for the pion mass, no attempt has been made to
extrapolate the parameters to physical values of the quark mass.}
\end{table}

\begin{table}
\begin{tabular}{ccrrrr}
                &                    & streamline         & quenched
                & ratio ansatz       & RILM        \\  \tableline
$m_N$           & $[\rm GeV]$        &    1.019           &    1.013
                & 0.983              &    1.040    \\
$\lambda_N^1$   & $[{\rm GeV}^3]$    &    0.026           &    0.029
                & 0.021              &    0.037    \\
$\lambda_N^2$   & $[{\rm GeV}^3]$    &    0.061           &    0.074
                & 0.048              &    0.093    \\
$m_\Delta$      & $[\rm GeV]$        &    1.428           &    1.628
                & 1.372              &    1.584    \\
$\lambda_\Delta$& $[{\rm GeV}^3]$    &    0.027           &    0.040
                & 0.026              &    0.036    \\
\end{tabular}
\caption{Baryon parameters in the different instanton ensembles. All
quantities are given in units of GeV. The current quark mass is $m_u
=m_d=0.1\Lambda$. }
\end{table}

\begin{table}
\begin{tabular}{crrrr}
correlator          &  definition       & resonance contribution &
$SU(2)_A$           &  $U(1)_A$    \\ \tableline
$\Pi_1^N(x)$        & $<{\rm} tr(\eta_1(x)\bar\eta_1(0))>$
                    & $|\lambda_1^{N^+}|^2m_{N^+}
                      -|\lambda_1^{N^-}|^2m_{N^-}$
                    & no  & no \\
$\Pi_2^N(x)$        & $<{\rm} tr(\gamma\cdot\hat x\eta_1(x)\bar\eta_1(0))>$
                    & $|\lambda_1^{N^+}|^2+|\lambda_1^{N^-}|^2$
                    & yes  & yes \\
$\Pi_3^N(x)$        & $<{\rm} tr(\eta_2(x)\bar\eta_2(0))>$
                    & $|\lambda_2^{N^+}|^2m_{N^+}
                      -|\lambda_2^{N^-}|^2m_{N^-}$
                    & no  & no \\
$\Pi_4^N(x)$        & $<{\rm} tr(\gamma\cdot\hat x\eta_2(x)\bar\eta_2(0))>$
                    & $|\lambda_2^{N^+}|^2+|\lambda_2^{N^-}|^2$
                    & yes  & yes \\
$\Pi_5^N(x)$        & $<{\rm} tr(\eta_1(x)\bar\eta_2(0))>$
                    & $\lambda_1^{N^+}
                       (\lambda_2^{N^+})^* m_{N^+}
                      -\lambda_1^{N^-}
                       (\lambda_2^{N^-})^* m_{N^-}$
                    & no  & no \\
$\Pi_6^N(x)$        & $<{\rm} tr(\gamma\cdot\hat x\eta_1(x)\bar\eta_2(0))>$
                    & $\lambda_1^{N^+}(\lambda_2^{N^+})^*
                      +\lambda_1^{N^-}(\lambda_2^{N^-})^*$
                    & yes  & no \\
\end{tabular}
\caption{Definition of various nucleon correlation functions. We also
give the form of the resonance contribution (with the propagators
suppressed) and the invariance properties under chiral $SU(2)_A$ and
$U(1)_A$ transformations.}
\end{table}

\pagestyle{empty}

\newpage
\begin{figure}
\begin{center}
\leavevmode
\epsfxsize=16cm
\epsffile{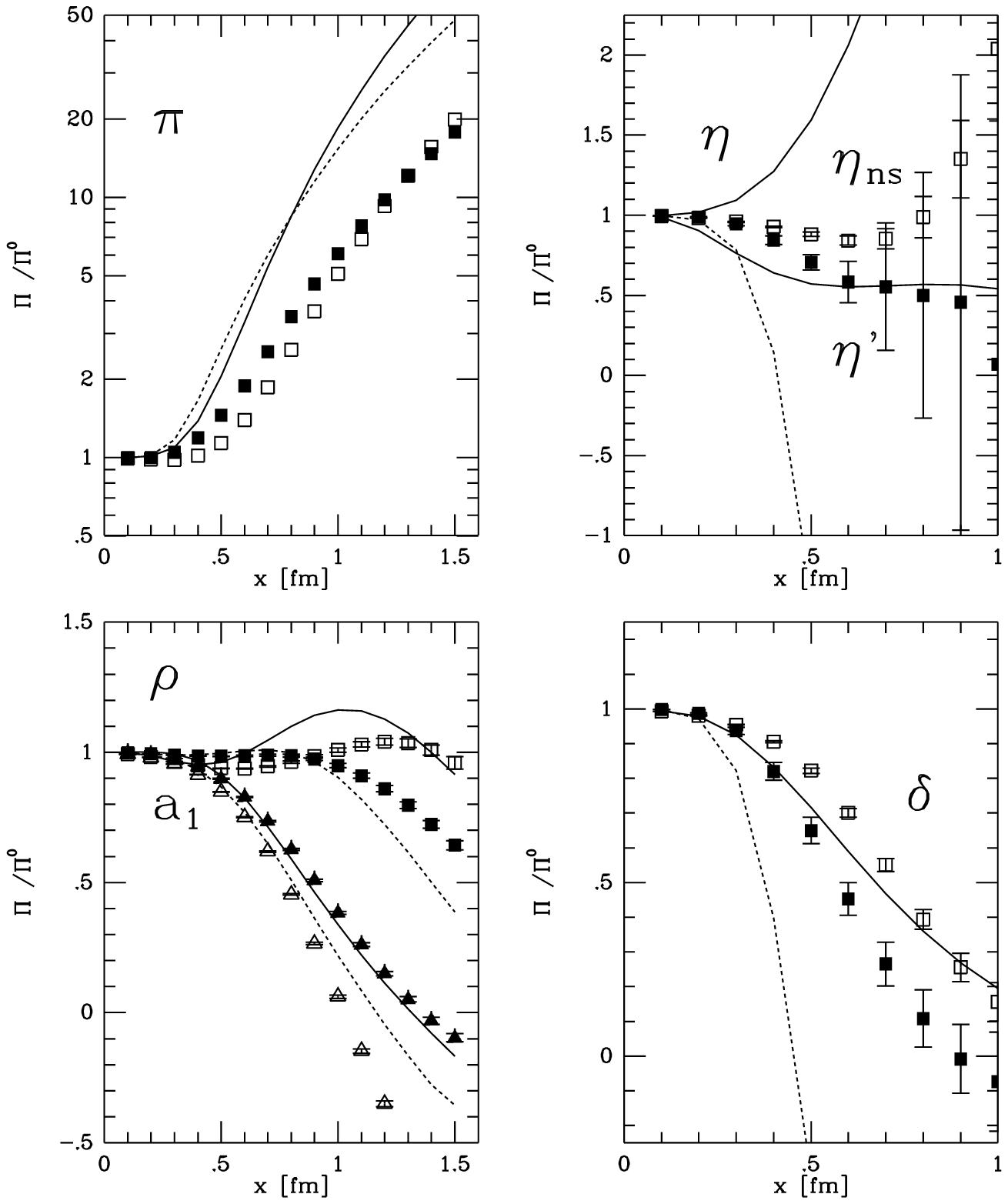}
\end{center}
\caption{}
\end{figure}
\vfill

\newpage
\begin{figure}
\begin{center}
\leavevmode
\epsfxsize=16cm
\epsfbox{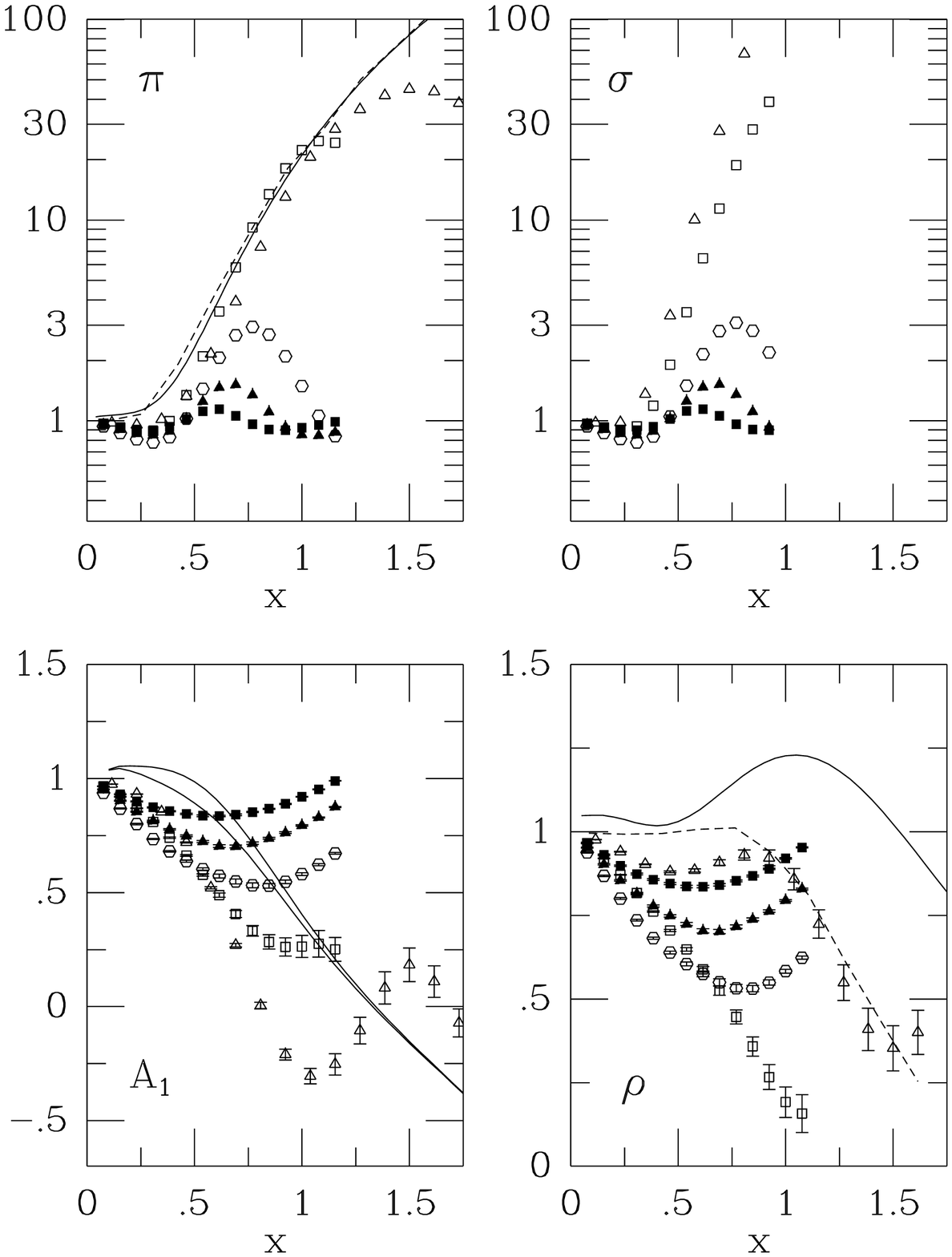}
\end{center}
\caption{}
\end{figure}
\vfill

\begin{figure}
\begin{center}
\leavevmode
\epsfxsize=16cm
\epsffile{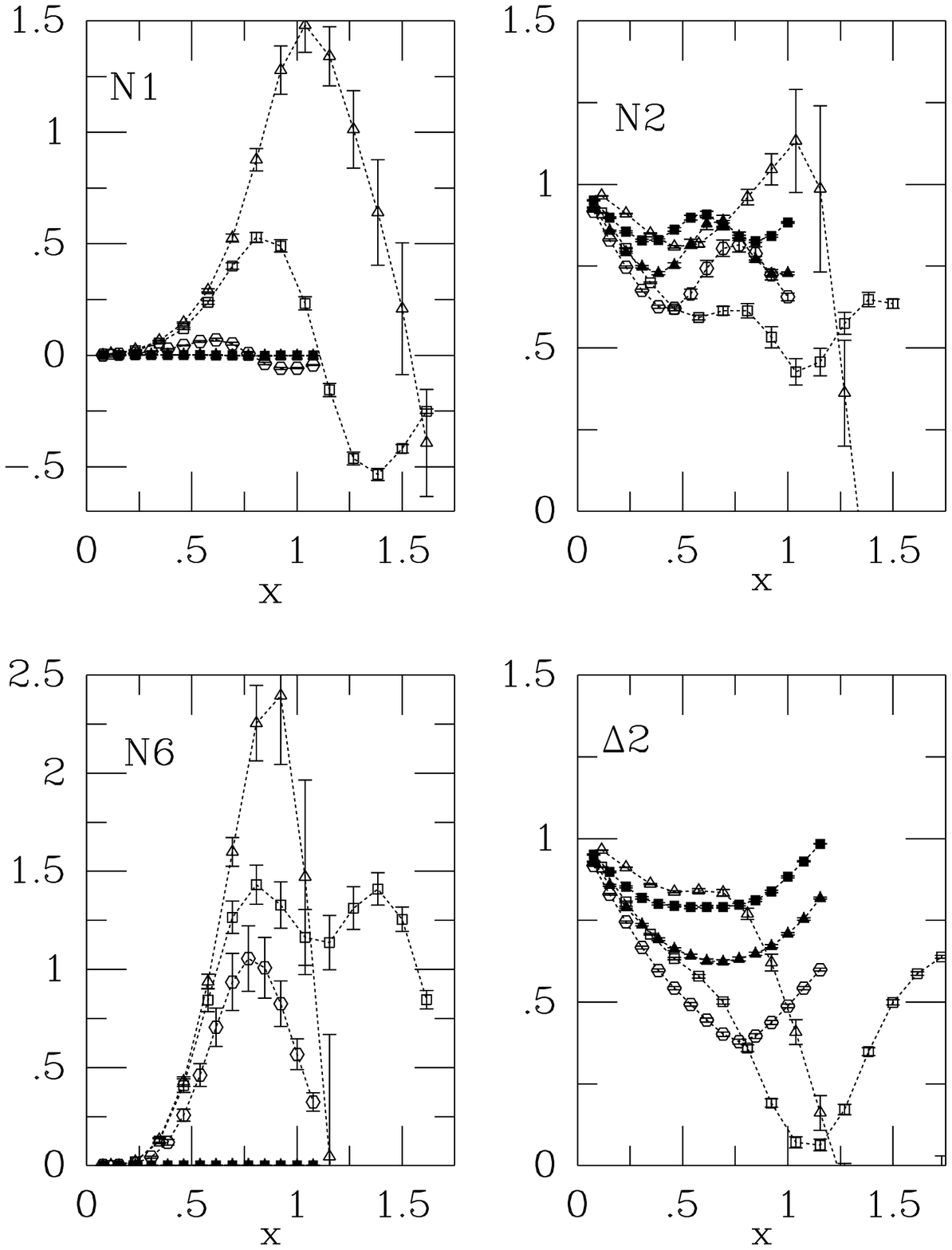}
\end{center}
\caption{}
\end{figure}
\vfill

\newpage
\begin{figure}
\begin{center}
\leavevmode
\epsfxsize=16cm
\epsffile{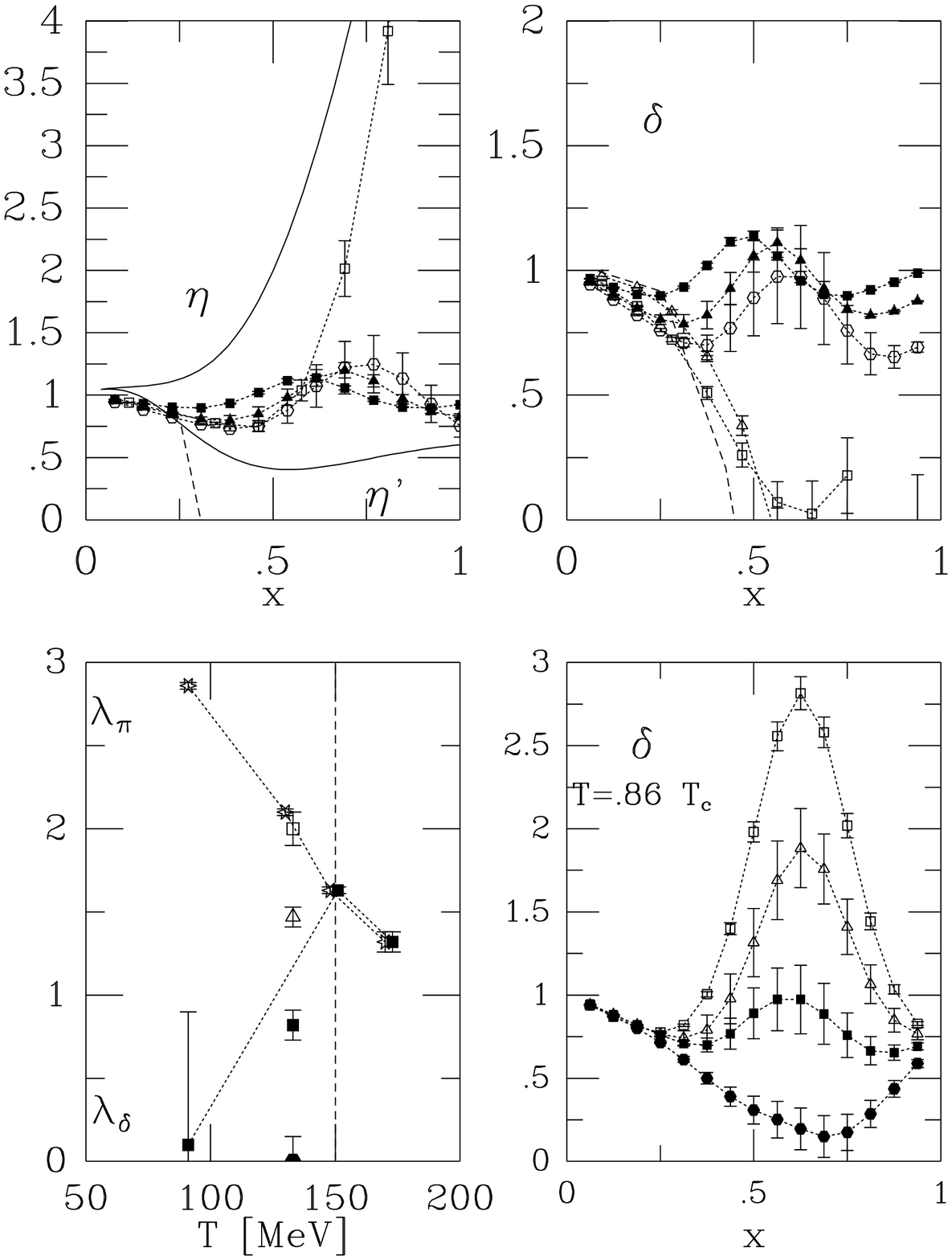}
\end{center}
\caption{}
\end{figure}
\vfill

\newpage
\begin{figure}
\begin{center}
\leavevmode
\epsfxsize=8cm
\epsffile{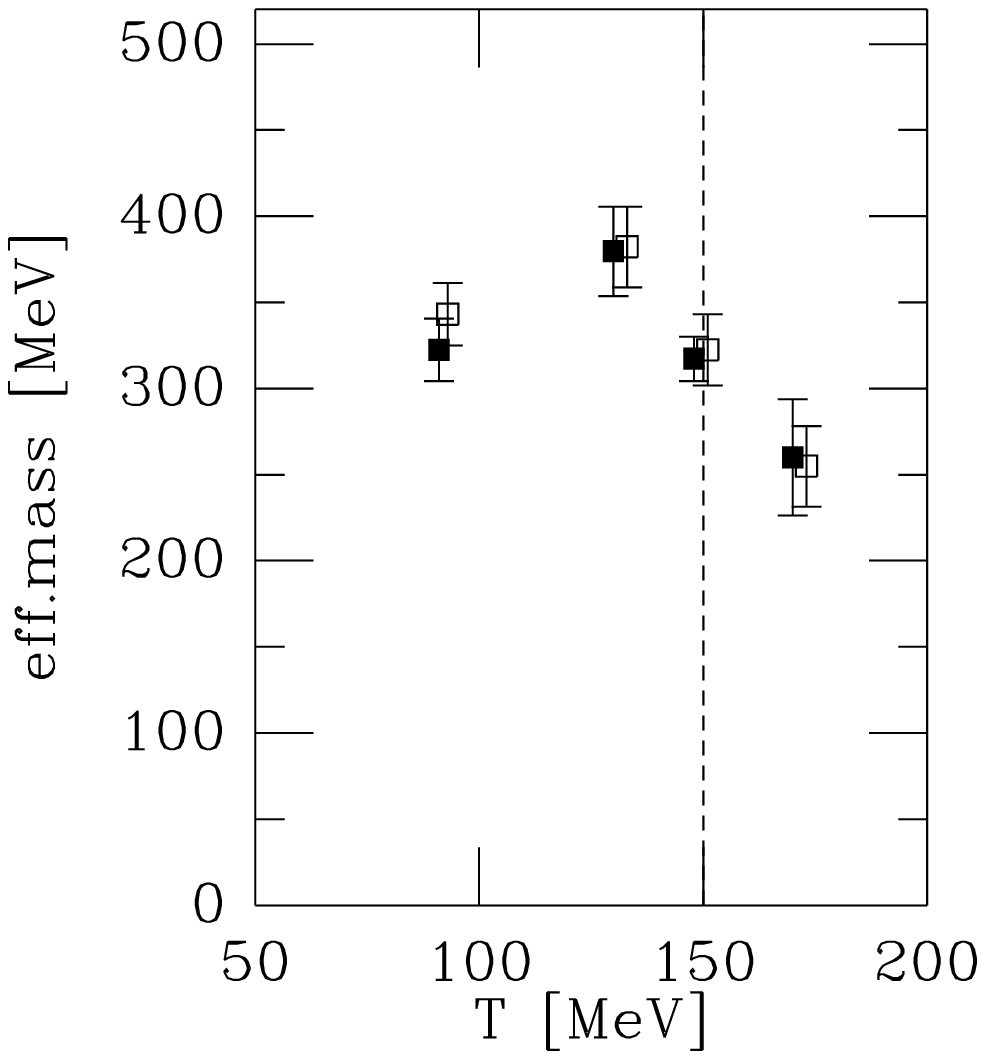}
\end{center}
\caption{}
\end{figure}
\vfill

\newpage
\begin{figure}
\begin{center}
\leavevmode
\epsfxsize=12cm
\epsffile{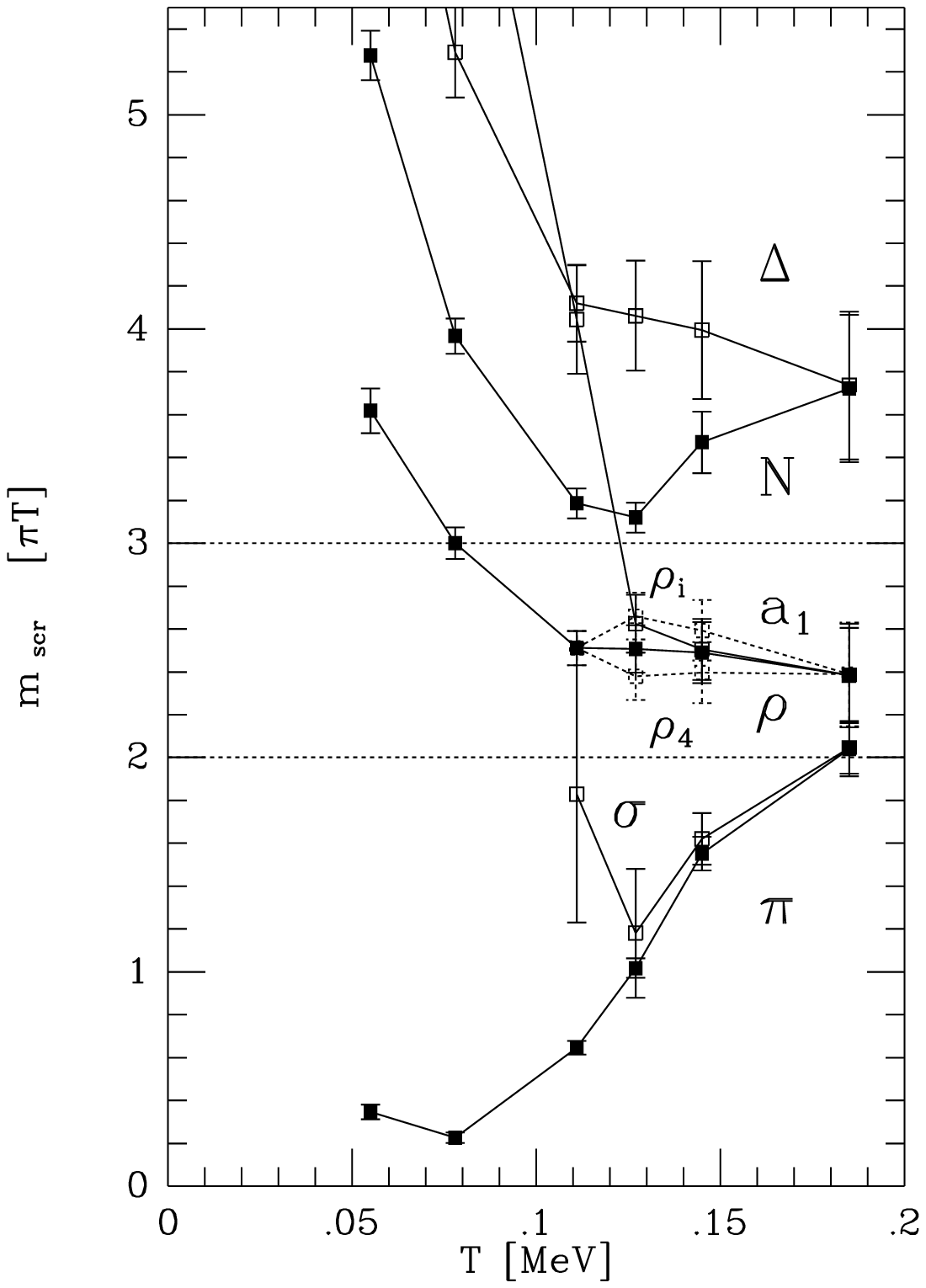}
\end{center}
\caption{}
\end{figure}
\vfill

\newpage
\begin{figure}
\begin{center}
\leavevmode
\epsfxsize=8cm
\epsffile{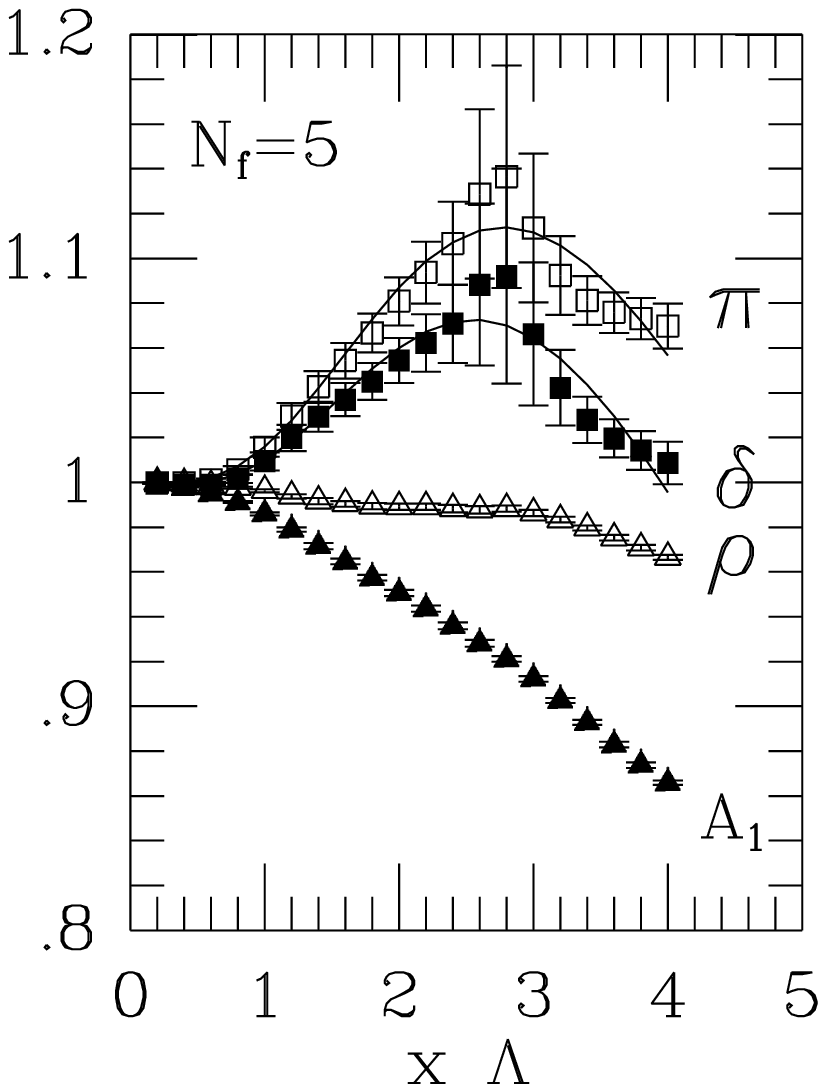}
\end{center}
\caption{}
\end{figure}
\vfill

\end{document}